\documentclass[aps,prd,superscriptaddress,preprint,tightenlines,nofootinbib,floatfix]{revtex4}

\usepackage{graphicx} 
\usepackage{dcolumn}  
\usepackage{bm}       
\usepackage{amsmath}
\usepackage{rotating}

\begin{document}

\preprint{CLNS 10/2064}  
\preprint{CLEO 10-02}    

\title{\boldmath Branching fractions for ${\chi_{cJ}\rightarrow p\bar{p}\pi^{0}}$, ${p\bar{p}\eta}$ and ${p\bar{p}\omega}$}

\author{P.~U.~E.~Onyisi}
\author{J.~L.~Rosner}
\affiliation{University of Chicago, Chicago, Illinois 60637, USA}
\author{J.~P.~Alexander}
\author{D.~G.~Cassel}
\author{S.~Das}
\author{R.~Ehrlich}
\author{L.~Fields}
\author{L.~Gibbons}
\author{S.~W.~Gray}
\author{D.~L.~Hartill}
\author{B.~K.~Heltsley}
\author{J.~M.~Hunt}
\author{D.~L.~Kreinick}
\author{V.~E.~Kuznetsov}
\author{J.~Ledoux}
\author{J.~R.~Patterson}
\author{D.~Peterson}
\author{D.~Riley}
\author{A.~Ryd}
\author{A.~J.~Sadoff}
\author{X.~Shi}
\author{W.~M.~Sun}
\affiliation{Cornell University, Ithaca, New York 14853, USA}
\author{J.~Yelton}
\affiliation{University of Florida, Gainesville, Florida 32611, USA}
\author{P.~Rubin}
\affiliation{George Mason University, Fairfax, Virginia 22030, USA}
\author{N.~Lowrey}
\author{S.~Mehrabyan}
\author{M.~Selen}
\author{J.~Wiss}
\affiliation{University of Illinois, Urbana-Champaign, Illinois 61801, USA}
\author{S.~Adams}
\author{M.~Kornicer}
\author{R.~E.~Mitchell}
\author{M.~R.~Shepherd}
\author{C.~M.~Tarbert}
\affiliation{Indiana University, Bloomington, Indiana 47405, USA }
\author{D.~Besson}
\affiliation{University of Kansas, Lawrence, Kansas 66045, USA}
\author{T.~K.~Pedlar}
\author{J.~Xavier}
\affiliation{Luther College, Decorah, Iowa 52101, USA}
\author{D.~Cronin-Hennessy}
\author{J.~Hietala}
\author{P.~Zweber}
\affiliation{University of Minnesota, Minneapolis, Minnesota 55455, USA}
\author{S.~Dobbs}
\author{Z.~Metreveli}
\author{K.~K.~Seth}
\author{T.~Xiao}
\author{A.~Tomaradze}
\affiliation{Northwestern University, Evanston, Illinois 60208, USA}
\author{S.~Brisbane}
\author{J.~Libby}
\author{L.~Martin}
\author{A.~Powell}
\author{P.~Spradlin}
\author{G.~Wilkinson}
\affiliation{University of Oxford, Oxford OX1 3RH, UK}
\author{H.~Mendez}
\affiliation{University of Puerto Rico, Mayaguez, Puerto Rico 00681}
\author{J.~Y.~Ge}
\author{D.~H.~Miller}
\author{I.~P.~J.~Shipsey}
\author{B.~Xin}
\affiliation{Purdue University, West Lafayette, Indiana 47907, USA}
\author{G.~S.~Adams}
\author{D.~Hu}
\author{B.~Moziak}
\author{J.~Napolitano}
\affiliation{Rensselaer Polytechnic Institute, Troy, New York 12180, USA}
\author{K.~M.~Ecklund}
\affiliation{Rice University, Houston, Texas 77005, USA}
\author{J.~Insler}
\author{H.~Muramatsu}
\author{C.~S.~Park}
\author{E.~H.~Thorndike}
\author{F.~Yang}
\affiliation{University of Rochester, Rochester, New York 14627, USA}
\author{S.~Ricciardi}
\affiliation{STFC Rutherford Appleton Laboratory, Chilton, Didcot, Oxfordshire, OX11 0QX, UK}
\author{C.~Thomas}
\affiliation{University of Oxford, Oxford OX1 3RH, UK}
\affiliation{STFC Rutherford Appleton Laboratory, Chilton, Didcot, Oxfordshire, OX11 0QX, UK}
\author{M.~Artuso}
\author{S.~Blusk}
\author{R.~Mountain}
\author{T.~Skwarnicki}
\author{S.~Stone}
\author{J.~C.~Wang}
\author{L.~M.~Zhang}
\affiliation{Syracuse University, Syracuse, New York 13244, USA}
\author{G.~Bonvicini}
\author{D.~Cinabro}
\author{A.~Lincoln}
\author{M.~J.~Smith}
\author{P.~Zhou}
\author{J.~Zhu}
\affiliation{Wayne State University, Detroit, Michigan 48202, USA}
\author{P.~Naik}
\author{J.~Rademacker}
\affiliation{University of Bristol, Bristol BS8 1TL, UK}
\author{D.~M.~Asner}
\author{K.~W.~Edwards}
\author{J.~Reed}
\author{K.~Randrianarivony}
\author{A.~N.~Robichaud}
\author{G.~Tatishvili}
\author{E.~J.~White}
\affiliation{Carleton University, Ottawa, Ontario, Canada K1S 5B6}
\author{R.~A.~Briere}
\author{H.~Vogel}
\affiliation{Carnegie Mellon University, Pittsburgh, Pennsylvania 15213, USA}
\collaboration{CLEO Collaboration}
\noaffiliation

\date{May 28, 2010}

\begin{abstract} 
Using a sample of 25.9 million $\psi(2S)$ decays acquired with the CLEO-c detector at the CESR $e^{+}e^{-}$ collider, we report branching fractions for the decays ${\chi_{cJ}\rightarrow p\bar{p}\pi^{0}}$, ${p\bar{p}\eta}$ and ${p\bar{p}\omega}$, with ${J=0,1,2}$.  Our results for ${\mathcal{B}(\chi_{cJ}\rightarrow p\bar{p}\pi^{0})}$ and ${\mathcal{B}(\chi_{cJ}\rightarrow p\bar{p}\eta)}$  are consistent with, but more precise than, previous measurements.  Furthermore, we include the first measurement of ${\mathcal{B}(\chi_{cJ}\rightarrow p\bar{p}\omega)}$.
\end{abstract}

\maketitle

Recent theoretical results~\cite{Barnes1,Barnes2,Barnes3} have highlighted the value of studying ${\Psi\rightarrow p\bar{p}M}$ hadronic decay processes, where $\Psi$ represents any $c\bar{c}$ bound state and $M$ is a light meson.  The application of these models allows measured ${\Psi\rightarrow p\bar{p}M}$ partial widths to be used to estimate the production cross sections for ${\sigma(p\bar{p}\rightarrow\Psi M)}$, circumventing the calculation of some of the complicated underlying QCD processes.  Calculations of this sort are interesting, for example, in the context of the future PANDA experiment~\cite{PANDALoI} which will exploit associated charmonium production in $p\bar{p}$ annihilation ${(p\bar{p}\rightarrow\Psi M)}$ in its search for exotic charmonia.  Since the values of $\Gamma(\Psi\to p\bar{p}M)$ serve as key inputs for these calculations, the same authors have also developed techniques for calculating $\Gamma(\Psi\to p\bar{p}M)$~\cite{Barnes4}, which can be tested with experimental data.  In their meson emission model, they assume the sequential decay ${\Psi\rightarrow p\bar{p}\rightarrow p\bar{p}M}$, and by applying techniques developed in~\cite{Barnes1,Barnes2,Barnes3}, they estimate ${\Gamma(\Psi\rightarrow p\bar{p}M)}$ using the measured ${\Psi\rightarrow p\bar{p}}$ widths and well-known $p\bar{p}M$ coupling constants.  If this sequential decay mechanism is in fact the dominant means by which ${\Psi\rightarrow p\bar{p}M}$ decays proceed, then the branching fractions to $p\bar{p}M$ final states would provide a means of extracting other meson-nucleon coupling constants~\cite{Barnes4}. 

This article describes measurements of the branching fractions for $\chi_{cJ}$ decays to three final states, $p\bar{p}\pi^{0}$, $p\bar{p}\eta$, and $p\bar{p}\omega$ using a sample of 25.9 million $\psi(2S)$ decays produced in $e^{+}e^{-}$ collisions at the Cornell Electron Storage Ring (CESR).  The first observations of $\chi_{cJ}$ decays to $p\bar{p}\pi^{0}$ and $p\bar{p}\eta$ were made by CLEO using a substantially smaller dataset of 3 million $\psi(2S)$ decays~\cite{CLEO2007}.  The $\psi(2S)$ produces a copious number of $\chi_{cJ}$ mesons via its  radiative $E1$ transitions, ${\psi(2S)\rightarrow\gamma\chi_{cJ}}$, with branching fractions of approximately 9\% for each of $J=0$, $1$, and $2$. We fully reconstruct decays of these secondary charmonia into $p\bar{p}\gamma\gamma$ and $p\bar{p}\pi^+\pi^-\pi^0$ final states using the CLEO-c apparatus.

The nearly hermetic CLEO-c~\cite{refCLEO} detector covers 93\% of the solid angle.  It features a 1~T superconducting solenoid housing drift chambers for tracking and charged particle identification and a ring imaging Cerenkov (RICH) system to further differentiate between charged particle species.  Also within the solenoid volume is an electromagnetic calorimeter composed of 7784 CsI(Tl) crystals.  The photon energy resolution in the calorimeter is 2.2\% at 1~GeV and 5\% at 100~MeV and the momentum resolution achieved using the drift chambers is typically 0.6\% at 1~GeV$/c$.    

In this analysis, we select events with either two or four charged tracks and at least three photons.  Candidate tracks are required to have momentum $p>18.4$~MeV$/c$ and originate within a 10~cm long, 2~cm radius cylindrical volume centered around the $e^+e^-$ interaction point. The $\pi^{\pm}$ candidate tracks are required to have specific ionization measurements ($dE/dx$) consistent with those expected for charged pions within 3 standard deviations.  The proton and anti-proton candidate tracks are required to have $dE/dx$ measurements within 4 standard deviations of the expected ionization losses for protons and anti-protons.  If RICH information is available for the event, it is used in conjunction with $dE/dx$ information to form joint likelihoods based on the hypothesis that the track is a proton, pion or kaon.  Candidate protons and anti-protons are then required to be more proton-like than pion-like or kaon-like.  Candidate photons are identified via the electromagnetic showers produced when incident on the calorimeter.  They must be associated with showers depositing more than $30$~MeV and have good separation from charged tracks.  Neutral pions and $\eta$ mesons are then reconstructed in their $\gamma\gamma$ decay modes. The invariant mass of each photon pair is calculated, and the pair is accepted as a neutral meson decay candidate when the invariant mass is within 3 standard deviations of the corresponding meson's rest mass. The detected four-momenta of all final state particles are improved via a series of kinematic fits.  In the $p\bar{p}\gamma\gamma$ and $p\bar{p}\pi^+\pi^-\pi^0$ modes, the four-momenta of the photon pairs in $\pi^0$ and $\eta$ candidates are constrained to the respective nominal rest masses taken from the Particle Data Group (PDG) report~\cite{PDG}.  All charged tracks are constrained to originate from a single event vertex that is within the beam spot, which is measured on a run-by-run basis by tracking the typical location of the event vertex. Finally, the $E1$ photon, $p$, $\bar{p}$ and $\pi^{0}$, $\eta$ or $\pi^+\pi^-\pi^0$ four-momenta are kinematically constrained to the initial state $\psi(2S)$ four momentum.  Events are then selected according to the $\chi^{2}/$d.o.f.\ of this four-constraint fit.  For ${p\bar{p}\gamma\gamma}$ final states, events are required to satisfy $\chi^{2}/$d.o.f.\ $< 5$  and for ${p\bar{p}\pi^{+}\pi^{-}\pi^{0}}$, $\chi^{2}/$d.o.f.\ $< 10$.  In instances where there is more than one possible combination of final state particles, $e.g.$, more than one pair of photons satisfy the $\pi^{0}$ selection criteria, the four constraint kinematic fit is performed for all possible permutations, and the combination with the lowest $\chi^{2}/$d.o.f.\ is selected.  This ambiguity in the final state particles occurs in around 10\% of events.
 
Further selection criteria to suppress backgrounds were investigated using a Monte Carlo (MC) sample of $1\times10^{8}$ $\psi(2S)$ decays generated using known partial widths from the PDG~\cite{PDG} and the models described in Ref.~\cite{MC} for any unknown branching fractions.  The dominant background in the $p\bar{p}\pi^0$ mode is from {$\psi(2S)\rightarrow\pi^0\pi^0J/\psi$}, {$J/\psi\rightarrow p\bar{p}$} events in which one photon from the two $\pi^0$ decays is soft enough not to skew the four-momentum to the extent that the event would fail the $\chi^2/$d.o.f.\ requirement.  To suppress these, we reject events when the invariant mass of the $p\bar{p}$ system is close to the $J/\psi$ mass, that is, when {$3.07<M(p\bar{p})<3.14$~GeV$/c^2$}.  The total remaining background does not peak in $M(p\bar{p}\pi^{0})$ near any of the $\chi_{cJ}$ masses and accounts for 6\% of the data passing our selection criteria.  There is another small background in the $p\bar{p}\eta[\gamma\gamma]$ mode which peaks in ${M(p\bar{p}\eta)}$ at the $\chi_{c2}$ mass from the process: ${\chi_{c2}\rightarrow\gamma J/\psi}$, ${J/\psi\rightarrow\gamma p\bar{p}}$.  The rate of this background is reduced by the initial requirement that $M(\gamma\gamma)$ is close to the $\eta$ rest mass, and it is further suppressed by rejecting events when the invariant mass of the $\gamma p\bar{p}$ system is close to the $J/\psi$ mass, $i.e.$, when {$3.07<M(\gamma p\bar{p})<3.12$~GeV$/c^2$}.  The remaining background accounts for 30\% of selected events but has no structure in $M(p\bar{p}\eta)$.  Additional requirements on kinematic variables are ineffective in suppressing background in the $p\bar{p}\pi^{+}\pi^{-}\pi^{0}$ final state since this is dominated by ${\chi_{cJ}\rightarrow p\bar{p}\pi^{+}\pi^{-}\pi^{0}}$ decays in which the three pions do not result from an $\eta$ or $\omega$.  These events conserve four-momentum and cannot be distinguished from resonant ${\left(\eta,\omega\right)\rightarrow\pi^{+}\pi^{-}\pi^{0}}$ decays on an event-by-event basis.  Instead, a term is included to account for this background in a fit to the $\eta$- and $\omega$-components of the decay.

The branching fractions for ${\chi_{cJ}\rightarrow p\bar{p}M}$ are calculated according to
\begin{equation}
\label{eqt:BF}
\begin{split}
\mathcal{B}\left(\chi_{cJ}\rightarrow p\bar{p}M\right) = &\frac{N_{M}}{ \epsilon_{M}N_{\psi(2S)}\mathcal{B}_{\gamma J}\mathcal{B}_Y}. 
\end{split}
\end{equation}
$N_{\psi(2S)}$ is the number of $\psi(2S)$ present in the data~\cite{Mendez}.  The signal efficiency of the combined CLEO-c apparatus, reconstruction and event selection algorithms, $\epsilon_{M}$, is evaluated via analysis of MC samples.   The branching fractions for ${\psi(2S)\rightarrow\gamma\chi_{cJ}}$, ${\mathcal{B}_{\gamma J}}$, are those measured by CLEO~\cite{CLEO2004}.  Values for ${\mathcal{B}_Y}$, which represents the branching fractions for ${M\rightarrow Y}$, where $Y$ represents either $\gamma\gamma$ or $\pi^{+}\pi^{-}\pi^{0}$, are taken from the  PDG~\cite{PDG}. The signal yield, $N_{M}$, is obtained via unbinned maximum likelihood fits to the data using slightly different techniques for the two distinct final states ${p\bar{p}\gamma\gamma}$ and ${p\bar{p}\pi^{+}\pi^{-}\pi^{0}}$. 

\begin{figure}
\includegraphics*[width=3.3in]{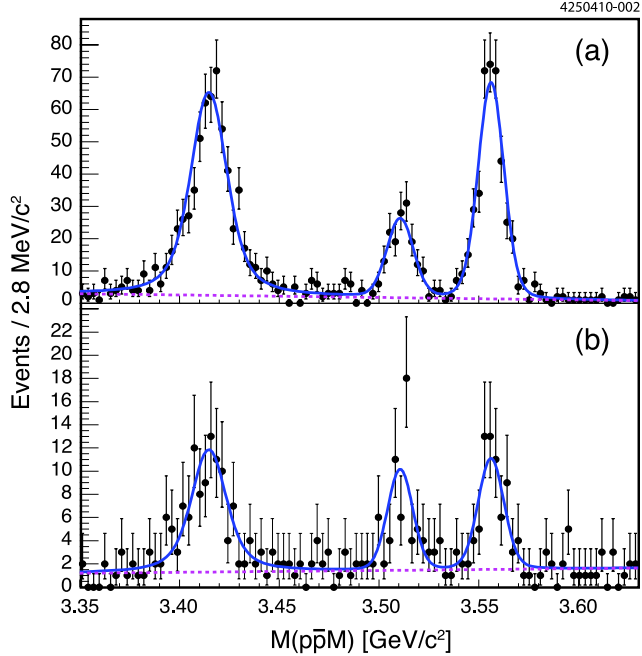}\label{fig:ppbarpi0fit}
\caption{Candidate $\chi_{cJ}$ mass spectrum for ${\chi_{cJ}\rightarrow p\bar{p}\pi^{0}}$ (a) and  ${\chi_{cJ}\rightarrow p\bar{p}\eta[\gamma\gamma]}$ (b).    Points with error bars are data, the solid lines show the fitted functions, and the dashed lines represent the linear background components of the fits.}
\label{fig:ppbarGGfit}
\end{figure}

In the first case, the yield is extracted via separate one-dimensional unbinned extended maximum likelihood fits to either the ${M(p\bar{p}\pi^{0})}$ or ${M(p\bar{p}\eta)}$ spectrum.  A linear background term is included in both fits to account for the small, flat background surviving our selection criteria.  The signal shapes are modeled by Breit-Wigner distributions convolved with Gaussian resolution functions.  The masses and widths of the Breit-Wigner distributions are fixed at the  PDG values~\cite{PDG} for the $\chi_{cJ}$.  The resolutions are fixed at values extracted from MC simulations.  The results of these fits are shown in Fig.~\ref{fig:ppbarGGfit}.

\begin{figure}
\includegraphics*[width=3.3in]{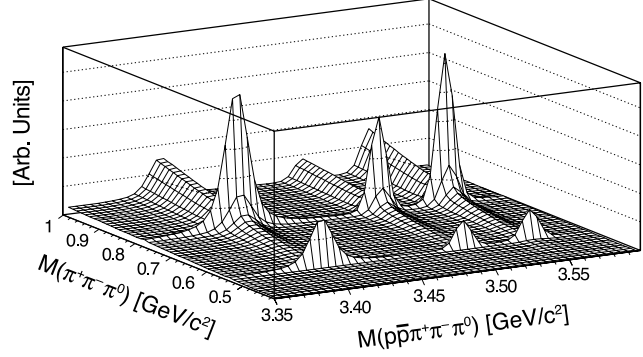}
\caption{The PDF that provides the best fit to the ${\chi_{cJ}\rightarrow p\bar{p}\pi^{+}\pi^{-}\pi^{0}}$ final state.  The six peaks due to the ${\chi_{cJ}\rightarrow p\bar{p}\eta}$ and ${\chi_{cJ}\rightarrow p\bar{p}\omega}$ signals are evident as well as the non-resonant and planar backgrounds.}
\label{fig:3piPDF}
\end{figure}

\begin{figure}
\includegraphics*[width=3.3in]{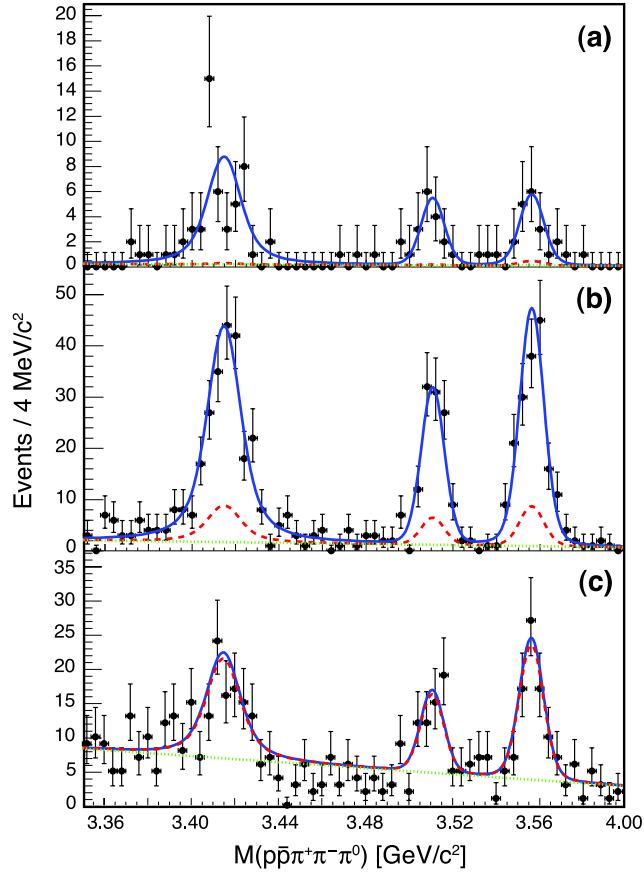}
\caption{Results of the two-dimensional fit to $M(\pi^{+}\pi^{-}\pi^{0})$ and $M(p\bar{p}\pi^{+}\pi^{-}\pi^{0})$ for the ${\chi_{cJ}\rightarrow p\bar{p}\pi^{+}\pi^{-}\pi^{0}}$ channel projected onto the $\chi_{cJ}$ candidate mass axis for three $M(\pi^+\pi^-\pi^0)$ regions: (a) 0.541-0.553~GeV$/c^2$, (b) 0.770-0.794~GeV$/c^2$, (c) 0.560-0.730~GeV$/c^2$.  Points with error bars are data, the solid lines are projections of the total fitted function, the dashed lines are the summed background components of the fit, and the dotted lines show the planar background components.}\label{fig:3pifitsa}
\end{figure}

\begin{figure}
\includegraphics*[width=3.3in]{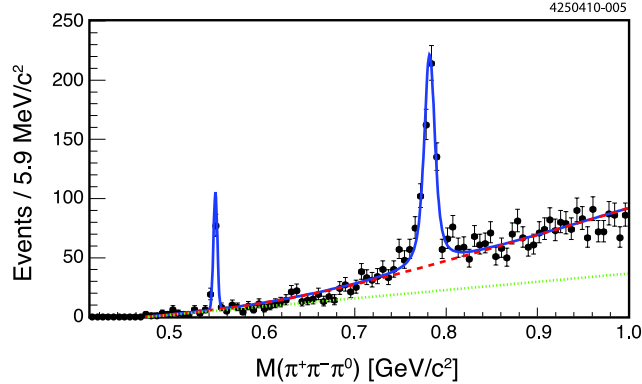}
\caption{Results of the two-dimensional fit to $M(\pi^{+}\pi^{-}\pi^{0})$ and $M(p\bar{p}\pi^{+}\pi^{-}\pi^{0})$ for the ${\chi_{cJ}\rightarrow p\bar{p}\pi^{+}\pi^{-}\pi^{0}}$ channel projected over the entire range of $M(p\bar{p}\pi^+\pi^-\pi^0)$ onto the $M(\pi^+\pi^-\pi^0)$ axis.  Points with error bars are data, the solid line is a projection of the total fitted PDF, the dashed line is the summed background components of the fit, and the dotted line shows the planar background component.}\label{fig:3pifitsb}
\end{figure}

The large non-$\eta$, non-$\omega$ background in the ${p\bar{p}\pi^{+}\pi^{-}\pi^{0}}$ channel led us to choose a signal extraction method consisting of a two-dimensional unbinned extended maximum likelihood fit in ${M(p\bar{p}\pi^{+}\pi^{-}\pi^{0})}$ and ${M(\pi^{+}\pi^{-}\pi^{0})}$ to simultaneously extract all six desired yields.  Fitting in both variables provides sensitivity to the non-resonant background shape over a wide range of ${M(\pi^{+}\pi^{-}\pi^{0})}$, which allows for a precise determination of the contribution in the $\eta$ and $\omega$ signal regions.   The $p\bar{p}\omega$ signal shapes are modeled as the product of Breit-Wigner distributions centered at the $\chi_{cJ}$ masses convolved with Gaussian resolution functions in ${M(p\bar{p}\pi^{+}\pi^{-}\pi^{0})}$ and a Breit-Wigner function centered at the $\omega$-mass and convolved with a Gaussian in ${M(\pi^{+}\pi^{-}\pi^{0})}$.  Similarly, the $p\bar{p}\eta$ signals are modeled as the product of Breit-Wigner distributions convolved with Gaussian resolution functions in ${M(p\bar{p}\pi^{+}\pi^{-}\pi^{0})}$ and, since the $\eta$ is sufficiently narrow, a Gaussian in ${M(\pi^{+}\pi^{-}\pi^{0})}$.  The non-resonant background is represented by a reversed ARGUS function~\cite{ARGUS} in ${M(\pi^{+}\pi^{-}\pi^{0})}$ multiplied by the convolution of Breit-Wigner functions and Gaussian resolution functions in ${M(p\bar{p}\pi^{+}\pi^{-}\pi^{0})}$. The ARGUS function threshold is fixed at the three pion mass threshold.  The non-peaking background can be well-described by a function linear in both ${M(p\bar{p}\pi^{+}\pi^{-}\pi^{0})}$ and ${M(\pi^{+}\pi^{-}\pi^{0})}$.  The masses and widths of the Breit-Wigner distributions are fixed to the PDG values~\cite{PDG}.  The Gaussian resolutions are allowed to float, although this parameter was constrained to be the same for each of the $\chi_{cJ}$'s.  The probability density function (PDF) that provides the best fit to data is shown in Fig.~\ref{fig:3piPDF}, and projections of the data and fit onto the ${M(p\bar{p}\pi^{+}\pi^{-}\pi^{0})}$ and ${M(\pi^{+}\pi^{-}\pi^{0})}$ axes are shown in Figs.~\ref{fig:3pifitsa} and \ref{fig:3pifitsb}.


The branching fractions are assigned systematic errors due to uncertainties in charged particle tracking efficiency (1\% per track), uncertainties in photon finding efficiency (2\% per photon), the uncertainty in the number of $\psi(2S)$ (2\%)~\cite{Mendez} and an error propagated from the uncertainty in the ${\psi(2S)\rightarrow \gamma\chi_{cJ}}$ branching fractions (5\%-7\%)~\cite{CLEO2004}.  The $\chi^{2}/$d.o.f.\ requirement introduces a small uncertainty (0\%-3\%) in the selection efficiency which is assessed by varying the requirement and repeating the analysis.  The error introduced from the choice of fitting technique is estimated by varying the fitted PDF and observing the change in efficiency-corrected yield.  Specifically, for the $p\bar{p}\gamma\gamma$ final states, the $\chi_{cJ}$ masses are varied by one standard deviation from the nominal PDG values~\cite{PDG} and the systematic uncertainty assigned as the average variation in observed yield between the two extremes of $\chi_{cJ}$ mass.  A similar variation is repeated for the $\chi_{cJ}$ widths.  The linear background term is replaced with a second order polynomial and the error assigned as the change in observed yield.  This uncertainty associated with the parametrization of the background shape is the dominant contribution to the fitting systematic and is largest in the ${p\bar{p}\eta}$ final state (1\%-8\%). The experimental resolution extracted from MC events is varied by $\pm25\%$ and the error assigned as the average yield for the two extremes.  The errors from each of these individual variations are added in quadrature to obtain the total error associated with the fitting technique.  

A similar set of variations is applied to the two-dimensional fits to the $p\bar{p}\pi^{+}\pi^{-}\pi^{0}$ final state. The $\chi_{cJ}$ masses and widths are varied by one standard deviation from the PDG values~\cite{PDG}.  This is repeated for the mass and width of the $\omega$ and the mass of the $\eta$.  In addition, the ARGUS function is replaced with a third order polynomial and the change in observed yield assigned as the uncertainty.  Finally, the range of $M(\pi^{+}\pi^{-}\pi^{0})$ fitted is varied from [0.41,1.0] to [0.53,0.85] and the uncertainty taken as the change in observed yield.  Again, the individual errors are summed in quadrature to obtain the total error associated with the fitting technique (1\%-5\%).

\begin{figure}
\includegraphics*[width=3.3in]{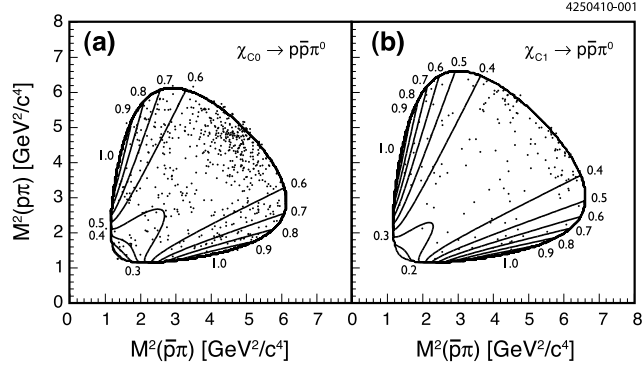}\label{fig:ppi0Dalitz0}
\caption{Dalitz plots for the $p\bar{p}\pi^{0}$ final state in decays of (a) ${\chi_{c0}}$ and (b) ${\chi_{c1}}$.  The contours indicate the density distributions predicted by the meson emission model~\cite{Barnes4}. \label{fig:ppi0Dalitz}}
\end{figure}

The efficiencies are extracted from MC simulations in which the ${\chi_{cJ}\rightarrow p\bar{p}M}$ decays populate phase space uniformly; $\epsilon_M$ is obtained by fitting either the 1D $M(p\bar{p}M)$ distribution or the 2D $M(p\bar{p}M)$, $M(p\bar{p}\pi^{+}\pi^{-}\pi^{0})$ distribution.  As a result, $\epsilon_M$ is an average of the efficiency over the $p\bar{p}M$ Dalitz plot.  We have investigated both how the data populate the Dalitz plot and how the efficiency varies across the Dalitz plot. Initially, $\chi_{cJ}$ candidates are selected in data via a requirement on $M(p\bar{p}M)$, and the surviving events are binned in terms of $M^{2}(pM)$ and $M^{2}(\bar{p}M)$.  The unbinned Dalitz plot is shown in Fig.~\ref{fig:ppi0Dalitz} for ${\chi_{c0,1}\rightarrow p\bar{p}\pi^{0}}$.  It is clear that the data do not populate phase space uniformly. In particular, a broad structure close to $p\bar{p}$ threshold is evident in ${\chi_{c0}\rightarrow p\bar{p}\pi^{0}}$.  The same selection criteria are then applied to the signal MC, and again the MC sample is binned in terms of $M^{2}(pM)$ and $M^{2}(\bar{p}M)$. By dividing out the number of generated MC events in each bin, we obtain the efficiency as a function of $M^{2}(pM)$ and $M^{2}(\bar{p}M)$.  To obtain the efficiency as a function of $M^{2}(pM)$ and $M^{2}(\bar{p}M)$ for the ${\chi_{cJ}\rightarrow p\bar{p}\omega}$ channels, a sideband subtraction in $M(\pi^+\pi^-\pi^0)$ is also required to suppress the non-resonant background (see Fig.~\ref{fig:3pifitsb}).  We find that the efficiencies are smoothly varying in $M^{2}(pM)$ and $M^{2}(\bar{p}M)$ in all channels with the exception of the region of $M(p\bar{p})$ close to the $J/\psi$ mass in the ${\chi_{cJ}\rightarrow p\bar{p}\pi^{0}}$ channels.  In this region the efficiency is considerably lower as a result of the requirement on $M(p\bar{p})$ needed to suppress backgrounds.

Variations in detection efficiency across the Dalitz plot together with resonant structures in the data could potentially lead to a large systematic uncertainty in the value of $\epsilon_M$.  We quantify this uncertainty by calculating the efficiency corrected yield in two ways.  First, the Dalitz plot for the data are integrated to obtain the yield and this is corrected using the efficiency averaged over $M^{2}(pM)$ and $M^{2}(\bar{p}M)$, a procedure that gives an efficiency-corrected yield close to that used in the nominal analysis.  Next, the data are corrected for efficiency as a function of $M^{2}(pM)$ and $M^{2}(\bar{p}M)$. The systematic uncertainty is then assigned as the difference in efficiency corrected yields obtained using these two methods and is less than or equal to 5\% in all channels excluding ${\chi_{c0}\rightarrow p\bar{p}\eta[\pi^+\pi^-\pi^0]}$.  In that case the uncertainty is slightly higher at 9\%.

The final product branching fractions, ${{\mathcal{B}_{\gamma J}} \times {\mathcal{B}(\chi_{cJ}\rightarrow p\bar{p}M)}}$, are given in Table~\ref{tab:final} where the errors shown are statistical then systematic.  The final branching fractions for the processes {$\chi_{cJ}\rightarrow p\bar{p}\pi^0$}, {$p\bar{p}\eta$} and {$p\bar{p}\omega$} are given in Table~\ref{tab:5}.  In this case the errors are statistical, systematic due to detector and analysis uncertainties, and a separate systematic error due to the uncertainty in the ${\psi(2S)\rightarrow\gamma\chi_{cJ}}$ branching fractions.  A weighted average of the two separate ${\mathcal{B}(\chi_{cJ}\rightarrow p\bar{p}\eta)}$ measurements is made taking into account correlated systematic errors.  The results are in good agreement with the previously published CLEO data~\cite{CLEO2007} with, as expected, a factor of around 3 smaller statistical errors.  

The meson emission model predictions for the two branching fractions ${\mathcal{B}(\chi_{c0}\rightarrow p\bar{p}\pi^{0})_\mathrm{theory} = 2.5\times10^{-4}}$ and ${\mathcal{B}(\chi_{c1}\rightarrow p\bar{p}\pi^{0})_\mathrm{theory} = 0.2\times10^{-4}}$~\cite{Barnes4} are well below our observed branching fractions, by factors of about 3 and 10 respectively. This suggests that meson emission, as described by this model, is not the dominant decay mechanism. This can be further demonstrated by comparing the theoretical Dalitz plot event densities calculated in reference~\cite{Barnes4} with our data; this comparison is shown in Fig.~\ref{fig:ppi0Dalitz}. The meson emission model predicts strength in regions of low $p\pi^{0}$ and $\bar{p}\pi^{0}$ invariant mass, whereas the data show a clear enhancement at low $p\bar{p}$ invariant mass.

In summary, we have presented new measurements of the branching fractions ${\mathcal{B}(\chi_{cJ}\rightarrow p\bar{p}\pi^{0})}$, ${\mathcal{B}(\chi_{cJ}\rightarrow p\bar{p}\eta)}$ and ${\mathcal{B}(\chi_{cJ}\rightarrow p\bar{p}\omega)}$.  We find good agreement with the available previous experimental results for ${\chi_{cJ}\rightarrow p\bar{p}\pi^{0}}$ and ${\chi_{cJ}\rightarrow p\bar{p}\eta}$, and the large CLEO data set allows us for the first time to disentangle the ${\chi_{cJ}\rightarrow p\bar{p}\omega}$ strength from the large non-resonant background.  Finally, we make a comparison with the meson emission model calculations of Barnes \textit{et~al.} and find that the sequential emission process described by the authors does not describe our data.

\begin{sidewaystable}[]
\caption{The product branching fractions $\mathcal{B} = \mathcal{B}_{\gamma J}\times\mathcal{B}(\chi_{cJ}\rightarrow p\bar{p}M)$\label{tab:final}.  Uncertainties are statistical, then systematic.  $\epsilon_M$ is the signal efficiency (described in the text).  Yield is the number of signal events evaluated via an unbinned maximum likelihood fit (described in the text). }
\medskip
\begin{center}
\begin{ruledtabular}
\begin{tabular}{lccc|ccc|ccc}
			& 	\multicolumn{3}{c}{$\chi_{c0}$}		& \multicolumn{3}{c}{$\chi_{c1}$}		& \multicolumn{3}{c}{$\chi_{c2}$}\\ 
 			& $\epsilon_M$(\%) & Yield				& {$\mathcal{B}(10^{-5})$}  	& $\epsilon_M$(\%)  & Yield				& {$\mathcal{B}(10^{-5})$}	& $\epsilon_M$(\%)  & Yield			& {$\mathcal{B}(10^{-5})$}\\ \hline
$p\bar{p}\pi^{0}$ 		   	&33.4 &610.9	&$7.15 \pm 0.34 \pm 0.47$ &36.1 &146.9	&$1.59 \pm 0.15 \pm 0.12$     &35.2 &406.6	&$4.51 \pm 0.24 \pm 0.33$ \\ 
$p\bar{p}\eta [\gamma\gamma]$ 		&30.6 &99.4 	&$3.18 \pm 0.42 \pm 0.23$ &33.1 &49.9 	&$1.48 \pm 0.25 \pm 0.17$     &33.1 &58.1 	&$1.72 \pm 0.27 \pm 0.16$ \\ 
$p\bar{p}\eta [\pi^+\pi^-\pi^0]$ 	&21.3 &51.9 	&$4.15 \pm 0.61 \pm 0.47$ &23.4 &18.1 	&$1.33 \pm 0.33 \pm 0.12$     &22.5 &19.9 	&$1.51 \pm 0.37 \pm 0.12$ \\ 
$p\bar{p}\omega$			&22.2 &263.0 	&$3.43 \pm 0.35 \pm 0.26$ &23.9 &113.7 	&$1.42 \pm 0.20 \pm 0.13$     &23.4 &185.5 	&$1.64 \pm 0.22 \pm 0.13$ \\ 
\end{tabular}
\end{ruledtabular}
\end{center}
\end{sidewaystable}
\begin{sidewaystable}[ht]
\caption{Final ${\chi_{cJ}\rightarrow p\bar{p}M}$ branching fractions\label{tab:5}.  Uncertainties are statistical, then systematic and then a separate systematic error due to the uncertainty in the ${\psi(2S)\rightarrow\gamma\chi_{cJ}}$ branching fractions. }
\medskip
\begin{center}
\begin{ruledtabular}
\begin{tabular}{lccc}
			& \multicolumn{3}{c}{${\mathcal{B}\left(\chi_{cJ}\rightarrow p\bar{p}M\right)}(10^{-4})$}\\ 
			& $J=0$	& $J=1$	& $J=2$\\ \hline
$p\bar{p}\pi^{0}$ 			&$ {7.76 \pm 0.37 \pm 0.51 \pm 0.39}$  &${1.75 \pm 0.16 \pm 0.13 \pm 0.11}$ &${4.83 \pm 0.25 \pm 0.35 \pm 0.31}$ \\ 
$p\bar{p}\eta [\text{mean}]$ 		&$ {3.73 \pm 0.38 \pm 0.28 \pm 0.19}$  &${1.56 \pm 0.22 \pm 0.14 \pm 0.10}$ &${1.76 \pm 0.23 \pm 0.14 \pm 0.11}$ \\ 
$p\bar{p}\omega$			&$ {5.57 \pm 0.48 \pm 0.42 \pm 0.28}$  &${2.28 \pm 0.28 \pm 0.16 \pm 0.14}$ &${3.68 \pm 0.35 \pm 0.26 \pm 0.24}$ \\
\end{tabular}
\end{ruledtabular}
\end{center}
\end{sidewaystable}

\begin{acknowledgments} 
The authors thank T.~Barnes for valuable discussions and providing theoretical results. We gratefully acknowledge the effort of the CESR staff 
in providing us with excellent luminosity and running conditions. 
D.~Cronin-Hennessy thanks the A.P.~Sloan Foundation. 
This work was supported by 
the National Science Foundation, 
the U.S. Department of Energy, 
the Natural Sciences and Engineering Research Council of Canada, and 
the U.K. Science and Technology Facilities Council. 
\end{acknowledgments} 

\end{document}